\documentclass[12pt]{article}                    
\usepackage{graphicx}
\usepackage{amssymb}
\usepackage{amsmath}
\newcommand{\bea}{\begin{eqnarray}}
\newcommand{\eea}{\end{eqnarray}}
\newcommand{\R}{\mathbb{R}}
\newcommand{\T}{\mathbb{T}}
\newcommand{\E}{\mathbb{E}}
\newcommand{\C}{\mathbb{C}}
\newcommand{\G}{\mathbf{G}}

\newcommand{\N}{\mathbb{N}}
\newcommand{\V}{\mathbb{V}}

\newcommand{\Velta}{\mbox{V}}

\newcommand{\sgn}{\mbox{sgn}}
\renewcommand{\bar}{\overline}

\renewcommand{\bar}{\overline}
\newcommand{\dZ}{Z^{\dagger}}

\DeclareMathOperator*{\pf}{\mbox{Pfaff}}
\newtheorem{thm}{Theorem}
\newtheorem{lemma}{Lemma}
\newtheorem{remark}{Remark}
\newtheorem{prop}{Proposition}
\begin{document}
\title{Averages of products of characteristic polynomials and the law of real
eigenvalues for the real Ginibre
ensemble.}

\author{Roger Tribe         \and
        Oleg Zaboronski}



\maketitle

\begin{abstract}
An elementary derivation of the Borodin-Sinclair-Forrester-Nagao
Pfaffian point process, which characterises the law of real eigenvalues for the
real Ginibre ensemble in the large matrix size limit, uses the averages of
products of characteristic polynomials.
This derivation reveals a number of interesting structures associated with 
the real Ginibre ensemble such as the hidden  symplectic symmetry of the statistics
of real eigenvalues and an integral representation for the $K$-point correlation function for any
$K\in \N$ in terms of 
an asymptotically exact integral over  the symmetric space $U(2K)/USp(2K)$.
%
%
\end{abstract}
\section{Introduction and main results}
\subsection{The real Ginibre ensemble of random matrices.}\label{intro}
The real Ginibre ensemble, denoted by GinOE$(N)$, is one of the classical random matrix ensembles
defined as the following probability measure on the space of real $N\times N$
matrices:
\bea
d\mu^{(N)}(M) = (\pi )^{- \frac{N^2}{2}} e^{ -  Tr M M^T}\prod_{i,j=1}^NdM_{ij},
\label{gpdf}
\eea 
where $\prod_{i,j=1}^NdM_{ij}$ is the Lebesgue measure on $\R^{N\times N}$.
This model was introduced in \cite{ginibre0} in 1965, but, unlike its 'self-adjoint' counterparts (the
models defined on spaces of symmetric, Hermitian and quaternionic self-dual matrices),
the calculation of the correlation functions for the real Ginibre model took much longer. For example,
the joint probability density of eigenvalues was derived
by Lehman and Sommers in \cite{lehmann} in 1991 only, and it took another decade for the calculation
of the correlation functions to be carried out by
by Forrester, Nagao \cite{forrester2007eigenvalue} and Borodin, Sinclair  \cite{borodin2009ginibre}. 
They independently discovered that the law of GinOE$(N)$ eigenvalues is a Pfaffian point process and determined
its kernel. 

The statistics of real eigenvalues for the real Ginibre ensemble turns out to be particularly interesting. 
It has been known since the work by Edelman, Kostlan and Shub
\cite{edelman1994many} that a large random Ginibre matrix has $O(\sqrt{N})$ eigenvalues. One of the results
of \cite{forrester2007eigenvalue}, \cite{borodin2009ginibre} is that the marginal law of real eigenvalues is also a Pfaffian
point process. It turns out, that the large-$N$ limit of this point process coincides (up to a Brownian rescaling) with the fixed-time law of annihilating
Brownian motions on the real line \cite{EJP942}. It is important to stress that, 
unlike the link between the statistics of GUE and Dyson Brownian motions, this does not extend to multi-time
statistics, see \cite{roger_oleg_2014}.  However, it does suggest that the
Pfaffian point process at hand maybe universal, where the corresponding
universality class contains both non-equilibrium interacting particle  systems and the non-symmetric
ensembles of random matrices.

We recall that the well-known results of Borodin, Sinclair, Forrester and Nagao concerning
the behaviour of real eigenvalues for the real Ginibre ensemble 
can be obtained without a reference to Lehmann-Sommers distribution. Instead, a duality relation
between GinOE$(N)$ and GinOE$(N-K)$ allows one to express a $K$ point correlation function for GinOE$(N)$
in terms of the expectation of the product of characteristic polynomials for GinOE$(N-K)$ \cite{roger_oleg_2014}. Here $N,K \in \N$, $K<N$.
The latter is easy to calculate using the Berezin calculus of anti-commuting variables.
Overall, the computation turns out to be surprisingly similar to the derivation
of the fixed time law for the annihilating Brownian motions carried out in \cite{masser2001method},
\cite{EJP942}
which relied on Markov duality between a finite and infinite systems of annihilating Brownian motions.
Roughly, the method is to 
"linearise" the model by finding sufficiently many expectation values which (i) determine the law of real eigenvalues;
(ii) can be characterised as solutions to a linear initial value problem, which are easy to write down 
explicitly . 

Our re-derivation of the Borodin-Sinclair-Forrester-Nagao Pfaffian point process 
reveals a couple of interesting mathematical structures associated with the real Ginibre ensemble in the large-$N$ limit:
firstly, the $K$-point correlation function of the real eigenvalues is given by the density of the eigenvalue
distribution for the Mehta-Pandey model interpolating between GUE$(K)$ and GSE$(K)$ at the
anti-self-dual point \cite{mehta1999some}, \cite{mehtapandey1983}.

Secondly, the Mehta-Pandey integral representing the $K$-point correlation function turns out to be asymptotically
exact, in the sense that the leading order term of its stationary phase approximation coincides
with an exact answer. It belongs to a novel family of asymptotically exact integrals over symmetric 
spaces generalising the celebrated Itzykson-Zuber integrals. 

The primary aim of the paper is to study the integral formulae for the multi-point
correlation functions for the real Ginibre ensemble using the heat kernel method and
the closely related proof of the asymptotic exactness of the integrals. To make the presentation
self-contained, we review the results of \cite{roger_oleg_2014} concerning the derivation
of the law of the real eigenvalues for the bulk scaling limit of GinOE$(N)$. 

The rest of the paper is organised as follows. In the rest of the introduction, we will recall the definition of
the random function counting the parity of the number of real eigenvalues in a semi-infinite interval, which
we refer to as 'spin variables' (subsection \ref{tools}). The expectations of the spin variables can be related
to the expectations of products of characteristic polynomials using the Householder 
transformation \cite{householder1958unitary}. We will then
state and discuss the main results (subsection \ref{mains}). The proofs are presented in section \ref{proofs}.
\subsection{The linearising 'spin' variables for the real Ginibre ensemble.}\label{tools}
The spin variable associated with a real valued matrix $M$ is the function $s(M):\R \rightarrow \{\pm 1\}$:
\bea
s_x(M) = (-1)^{\Lambda^M(-\infty,x)}, \quad x \in \R,
\label{spins}
\eea
where $\Lambda^M(a,b)$ is the number of {\it real} eigenvalues of $M$ in the interval $(a,b)\subset \R$.
Note an analogy between the spin variables (\ref{spins}) and spins in a 
one-dimensional spin chain with real
eigenvalues playing the role of domain walls. Spin variables are crucial in 
linearising the moment equations for
annihilating random walks and/or Brownian motions, see 
e.g. \cite{glauber1963time}, \cite{masser2001method}. 
We believe they will be useful for any random matrix model
with either purely real spectrum or such that the complex eigenvalues appear in conjugate pairs. Examples
include both the Hermitian matrix models such as GUE$(N)$ or GOE$(N)$, and the non-Hermitian ones
such as GinOE$(N)$.
 
The following elementary remark provides a tool for computing 
product moments of spin variables: the
spectrum of a real $N\times N$ matrix $M$
consists of real eigenvalues and pairs of conjugated complex eigenvalues. Therefore, 
\bea\label{new7}
s_x(M)=(-1)^{\# \left\{\mbox{ All eigenvalues of $M$ with real parts in $(-\infty,x)$}\right\}}.
\eea
Now, let us assume that $M\sim\mbox{GinO}E(N)$. 
As a pair of complex conjugated eigenvalues corresponds to a positive 
factor in the characteristic
polynomial, (\ref{new7}) implies that 
\bea\label{spin2det}
s_x(M)=\sgn\left(\det\left(M-xI\right)\right)=\frac{\det\left(M-xI\right)}{|\det\left(M-xI\right)|} \mbox{ a.s.}
\eea
Here we used that under the law of the real Ginibre ensemble the probability that $x$ is an eigenvalue of $M$
is zero.
We will recall that when computed with the help of Householder 
transformations \cite{householder1958unitary}, the product moments of spin variables
are expressed in terms of the product moments 
of characteristic polynomials for the real Ginibre ensemble of a smaller size. 
In the large-$N$ limit, these correlation functions will be shown to satisfy a linear parabolic
partial differential equation on the Weyl chamber, thus confirming the claimed
linearisation of the real Ginibre ensemble in terms of spin variables. Of course,
the statistics of characteristic polynomials can be studied directly
 using the Lehmann-Sommers distribution, see for example
 \cite{akemann2008characteristic}, \cite{sommers2009schur}.

All multi-point (Lebesgue) densities for real eigenvalues can be restored from the
moments of spin variables. Namely we have the following relation:
\bea
&& \hspace{-.4in} \rho^{(N)}(x_1, x_2,\ldots, x_K) \nonumber \\
& = & \left. \left(-\frac{1}{2}\right)^K
\left(\prod_{k=1}^K \frac{\partial}{\partial y_{k}}\right)
\mathbb{E}_N\left[\prod_{m=1}^{K} s_{x_m}\left(M\right)s_{x_m+y_m}\left(M\right) \right]
\right|_{y_m=0+, \, m=1,2\ldots , K}
\label{eq:dens}
\eea
where
$\E_N$ is the expectation with respect to GinOE$(N)$ and
$\rho^{(N)}(x_1, x_2,\ldots, x_K)$ is the correlation function of order $K$ (the factorial density of
order $K$) for the point process corresponding to the law of real eigenvalues of GinOE$(N)$,
see \cite{EJP942} for details of the rigorous derivation of (\ref{eq:dens}).
From a purely technical point of view, it is also useful to consider derivatives of moments of product spins
leading us to the so-called modified correlation functions defined as follows:
\bea\label{modense0}
\tilde{\rho}^{(N)}(x_1, x_2, \ldots, x_K):=\left(-\frac{1}{2}\right)^K 
\left(\prod_{m=1}^{K}\partial_{x_m}\right)\E_N\left[  \prod_{k=1}^K s_{x_k}(M)\right].
\eea
Equivalently, in terms of the counting measure $\Lambda^M$,
\bea\label{modense}
 \tilde{\rho}^{(N)}(x_1,x_2,\ldots, x_K) \prod_{k=1}^K dx_k
= \mathbb{E}_N \left[ \prod_{k=1}^K s_{x_k}(M)  \Lambda^{M}(dx_k) \right].
\label{rhom}
\eea
The above formula is an equality between measures acting on direct products of disjoint intervals (with $dx_k$ on the left hand side
being a standard abbreviation for the Lebesgue measure on $\R$).
The product moments of spin variables can be iteratively restored from the modified densities
by a $K$-dimensional integration:
\begin{eqnarray}\label{eq:sss1d}
\mathbb{E}_{N}\left[\prod_{k=1}^K \left(s_{x_k}(M)-1\right) \right] = (-2)^K\left(\prod_{k=1}^K \int_{-\infty}^{x_k} dy_k\right)
\tilde{\rho}^{(N)}(y_1,y_2,\ldots, y_{k}).
\end{eqnarray}

We are ready to state the main results of the paper. We use the convention that
$C$ denotes a constant, whose dependence will be indicated (for example $C_{K}$) but whose exact value
is unimportant and may change form line to line. For constant whose value we wish to record we use subscripts
for future reference (for example $c_1(K,N)$). 
\subsection{Results}\label{mains}
We first recall is a simple relation between the modified density expressed  in terms
of the product moment of spin variables
and moments of characteristic polynomials, which is valid for real Ginibre matrices of
any size $N\leq \infty$. 
It can be succinctly 
expressed by the following duality formula:
\begin{lemma}[\cite{roger_oleg_2014}]\label{lm1} Choose $-\infty<x_1<x_2<\ldots<x_{2K-1}<x_{2K}<\infty$. Then
\bea\label{res1}
&& \hspace{-.3in}\left(\prod_{k=1}^{2K} \partial_{x_k}\right) 
\mathbb{E}_{N}
\left[
\prod_{l=1}^{2K}
\frac{det(M-x_{l})}{|det(M-x_{l})|}\right]
\nonumber\\
&=&
c_1(K,N)e^{-\sum_{k=1}^Kx_k^2}\Velta(x)\mathbb{E}_{N-2K}
\left[\prod_{l=1}^{2K} det(M-x_l)\right],~2K<N\in \N,
\eea
where 
$c_1(K,N)>0$ is an explicit constant and $\Velta(x):=\prod_{i<j}^{2K}(x_j-x_i)$ is the Vandermonde determinant.
\end{lemma}
In other words, the derivatives with respect to the argument
s of characteristic polynomials
 appear to 'cancel' the denominators in the product of ratios of characteristic polynomials in the left hand side leaving
one with the expected value of the product of characteristic polynomials on the right hand side.
The latter are easy to evaluate in the large-$N$ limit using the supersymmetric formalism, see \cite{ss}
for a review. Thus the formula (\ref{res1}) leads to an integral representation for the 
modified $K$-point density of real eigenvalues in the large $N$ limit
\begin{equation} \label{modlimit}
\tilde{\rho}(x_1, x_2, \ldots, x_K):=
\lim_{N\rightarrow \infty} \tilde{\rho}^{(N)}(x_1, x_2, \ldots, x_K).
\end{equation}
\begin{thm}[\cite{roger_oleg_2014}]\label{thm1} (Ginibre ensemble and anti-self dual Gaussian symplectic ensembles.)
Let $K$ be an even natural number. Let $X=\mbox{Diag}(x)$ be a diagonal $K\times K$ matrix with the diagonal entries 
$x \in \R^K$ satisfying 
$x_1<x_2<\ldots <  x_K \in \R$. 
Then the limit (\ref{modlimit}) exists and is given by 
\bea\label{rho_mod}
\tilde{\rho}(x_1, x_2, \ldots, x_K)=C_K\Velta(x)  \int_{U(K)} \mu_H(dU)
e^{-\frac{1}{2}Tr\left(H-H^R\right)^2}
\eea
where $C_K$ is a positive constant, $H=UXU^\dagger$ is a Hermitian 
matrix , $\mu_H$ is Haar measure
on the unitary group $U(K)$,and  $H^R=JH^TJ$ is a symplectic involution of 
matrix $H$ using $J$ 
the canonical symplectic matrix.
\end{thm}
The integral on the right hand side of (\ref{rho_mod}) is a particular case of the elliptic
Gaussian matrix model which interpolates between the classical GUE and GSE ensembles. This model
was introduced and solved by Mehta and Pandey in \cite{mehta1999some}, \cite{mehtapandey1983}, see \cite{mehta2004random} for a review.
It is remarkable that it appears in the $N=\infty$ limit of the correlation function of characteristic polynomials for 
the real Ginibre ensemble, which does not have any apparent symplectic symmetry.
\begin{remark}
Let $\E_{MP(K)}$ be the expectation with respect to the anti-self-dual instance of $K\times K$ Mehta-Pandey model,
a matrix model defined on the space of $K\times K$ matrices by the measure 
$\exp{\left[-\frac{1}{2} Tr\left(H-H^R\right)^2 \right]}dH$,
where $dH$ is the Lebesgue measure. Then the statement of Theorem \ref{thm1} can be re-written
as a statement of matrix model duality as defined in e.g. \cite{Forrains}:
\bea
\lim_{N\rightarrow \infty} 
\left(\prod_{k=1}^K \partial_k\right)
 \E_{N}
\left[\prod_{\ell=1}^K\frac{det(M-x_\ell I)}{|det(M-x_\ell I)|}\right]
=C_K\E_{MP(K)}\left[\delta(\sigma_H-x)\right] \Velta(x),
\eea
where $\sigma_H$ is the spectrum of the self-adjoint matrix $H$.
\end{remark}
Mehta and Pandey use a Hubbard-Stratonovich transformation to reduce
the integral to the Itzykson-Zuber case.
In this paper we show that the integral in the right hand side of (\ref{rho_mod}) can be evaluated using the heat kernel method, 
the advantage of which is the possibility 
of a generalisation to an arbitrary symmetric space
of a compact Lie group $G$ with an involution.
Namely, we have the following statement proved in Section \ref{ss3}.
\begin{prop}\label{lm2} Define
\bea\label{intlm2}
I_t(X):=\int_{U(K)} \mu_H(dU)
e^{-\frac{1}{2t}Tr\left(H-H^R\right)^2},
\eea
where $H=UXU^\dagger$ and $X=\mbox{Diag}(x)$, and $x\in \R^K$ for even $K>0$.
Let $\tilde{\rho}_t(x)=C_K t^{-\frac{K(K+1)}{4}} \Velta(x)I_t(x)$ be a deformation
of $\tilde{\rho}$, which coincides with $\tilde{\rho}$ for $t=1$ and $x\in W^{(K)}:=
\{x: x_1<x_2<\ldots<x_K\}$.
Then $(\tilde{\rho}_t:t>0)$ is the unique distributional solution to the heat equation
\bea\label{ivp}
\left\{
\begin{array}{l}
(\partial_t-\frac{1}{8}\Delta)\tilde{\rho}_t(x)=0, t>0, \quad x\in \R^K\\
\tilde{\rho}_{0+}(x)=C_K \prod_{k=1}^{K/2}\delta^{\prime}(x_{2k}-x_{2k-1}), \quad x\in \R^K.
\end{array}
\right.
\eea
\end{prop}
The equation (\ref{ivp}) is easy to solve leading to a direct proof, avoiding marginalisation, of the following foundational result.
\begin{thm}[Borodin-Sinclair \cite{borodin2009ginibre}, Forrester-Nagao \cite{forrester2007eigenvalue}]\label{bsfn}
The bulk scaling limit of the law of real eigenvalues for GinOE$(N)$ is a Pfaffian point process:
\bea
\lim_{N\rightarrow \infty} \rho^{(N)}(x_1, x_2, \ldots, x_K)=\pf_{1\leq i,j\leq K} H(x_j-x_i),~K\geq 1,
\eea
where 
\bea
H(x)=\left( \begin{array}{cc}
-F''(x) & -F'(x) \\
F'(x) & sgn(x)F(|x|)   \end{array} \right),
\eea
and $F(x)=\pi^{-1/2}\int_{x}^\infty e^{-z^2} \,dz$.
\end{thm}
\begin{remark} 
This is Corollary 9 of \cite{borodin2009ginibre} and can also be easily 
restored from the results of \cite{forrester2007eigenvalue}.
\end{remark}
As a by-product of our proof of Theorem \ref{bsfn} one gets the 
following answer for the integral $I_t$: for all disjoint $(x_i)$
\bea\label{iz12}
I_t(x)&:=&\int_{U(K)} \mu_H(dU)
e^{-\frac{1}{2t}Tr\left(H-H^R\right)^2}\nonumber\\
&=&
C_K \frac{\pf_{1\leq i,j\leq K}\left[\frac{(x_i-x_j)}{\sqrt{t}}
e^{-\frac{(x_i-x_j)^2}{t}}\right]}{\Velta\left(\mathbf{\frac{x}{\sqrt{t}}}\right)}.
\eea
Our final result concerns the asymptotic exactness of the integral $I_{it}(x)$ for $t\in \R$,
 a result conjectured by Yan Fyodorov during an after-seminar discussion.
\begin{thm}\label{thm2}
The integral, for even $K>0$, 
\bea\label{itint}
I_{it}(x):=\int_{U(K)} \mu_H(dU)
e^{\frac{i}{2t}Tr\left(H-H^R\right)^2}
\eea
is asymptotically exact. In other words the exact expression for $I_{it}$ obtained from (\ref{iz12})
by replacing $t$ with $it$ coincides with the leading term of the stationary phase expansion, for small $t$,
of the integral in (\ref{itint}).
\end{thm}

\begin{remark}
At the moment, the precise reason for this localisation is unclear to us. 
In particular, the Duistermaat-Heckmann Theorem \cite{getzler} which is 
responsible for the exact localisation of the Itzykson-Zuber-Harish-Chandra
integral is not directly applicable to our case. Due to symplectic invariance 
of the integrand, the integral in  (\ref{iz12}) is taken over the symmetric 
space $U(K)/USp(K)$, where $USp(K)$ is the symplectic subgroup of $U(K)$. 
But $dimU(K)/USp(K)=K(K-1)/2$, which is even only if $K$ is divisible 
by $4$. So in general,   $U(K)/USp(K)$ is not even symplectic and the Duistermaat-Heckmann Theorem cannot apply. 
\end{remark}

The proof of 
Lemma \ref{lm1} can be found in Section \ref{ss1}; 
Theorem
\ref{thm1} is proved in Section \ref{ss2}; 
Proposition \ref{lm2} is proved in Section \ref{ss3}; Theorem \ref{bsfn} is proved in Section \ref{ss4};
Theorem  \ref{thm2} is proved in Section \ref{ss5}.
\section{Proofs.}\label{proofs}
\subsection{Lemma \ref{lm1}}
\label{ss1}
Following \cite{roger_oleg_2014},
let us fix $(x_1, x_2, \ldots, x_{2K})\in W^{(2K)}$.
We will assume $N$ to be even,
which helps us avoid tracking various $\pm$ signs. The proof for odd $N$ is similar.

It follows from (\ref{modense}) and (\ref{eq:sss1d}) that we need to evaluate
\bea\label{ginint}
\tilde{\rho}^{(N)} (x_1,x_2,\ldots,x_{2K}) \, \prod_{k=1}^{2K}dx_k =\frac{1}{4^K} \int_{R^{N^2}} dM \frac{e^{- Tr MM^T}}{\pi^{N^2/2}}
\prod_{k=1}^{2K} s_{x_k}(M)  \Lambda^{M}(dx_k).
\eea

The integral in the right hand side can be computed recursively using the Householder transform \cite{householder1958unitary}. 
The calculation is a direct generalisation of Edeleman's
calculation of the eigenvalue density for the real Ginibre ensemble \cite{edelman1994many}.

Let $M$ be a real $N \times N$ matrix
with a real eigenvalue $x$ and the corresponding eigenvector
$v \in S^+_{N-1}$, the upper half of the $N-1$ dimensional unit sphere in $\R^N$. Consider the following change of variables:
\bea\label{edt}
M = P_v M^e P_v
\eea
where $P_v$ is the Householder transformation \cite{householder1958unitary} 
that reflects in the hyperplane at right
angles to the vector $v-e_N$ (where $e_N$ is the unit vector $(0,\ldots,0,1)$), 
and $M^e$ is a block matrix
\bea\label{edt1}
M^e = \left( \begin{array}{cc}
M^e_0 & 0 \\
w^T & x
\end{array} \right)
\eea
with $M_0^e$ an $(N-1) \times (N-1)$ real matrix, $w \in R^{N-1}$ and $x \in \R$. 
The Jacobian of the Edelman's transformation
(\ref{edt}) is $|\det(M_0^e-xI)|$, see
\cite{edelman1994many}.

Let us perform the change of variables $M\rightarrow (M^e, v_{2K}, x_{2K})$ in the
integral (\ref{ginint}), where $x_{2K}$ is the eigenvalue of $M$ lying in $dx_{2K}$ and $v_{2K}$ is the corresponding eigenvector. 
Integrating over the half sphere $S_{N-1}^+$ and noticing the cancellation between the denominator
$|\det(M-x_{2K} I)|$ of the spin variable $S_{M}(x_{2K})$  and 
the Jacobian $|\det(M_0^e-x_{2K}  I)|$ of (\ref{edt}), we obtain
\begin{eqnarray}
 && \hspace{-.3in} 4^K\tilde{\rho}^{(N)}  (x_1,x_2,\ldots,x_{2K})  \, \prod_{k=1}^{2K-1}dx_k \\
   & = &   \frac12 |S_{N-1}| \pi^{-\frac{N-1}{2}} e^{-x_{2K}^2}
  \mathbb{E}_{N-1} \left[\det\left(M-x_{2K} I \right) \prod_{k=1}^{2K-1} s_{x_k}(M) \Lambda^{M}(dx_k)\right].\nonumber
\end{eqnarray}
A subsequent Edelman transform about the eigenvalue lying in $dx_{2K-1}$ yields
\begin{eqnarray}
 && \hspace{-.3in} 4^K\tilde{\rho}^{(N)}  (x_1,x_2,\ldots,x_{2K})  \, dx_1 \ldots dx_{K-2} \nonumber\\
& = &  \frac14 |S_{N-1}| |S_{N-2}|
\pi^{-\frac{N-1}{2}- \frac{N-2}{2}} e^{-x^2_{2K}-x^2_{2K-1}} (x_{2K-1}-x_{2K}) \\
&&  \hspace{.2in} \mathbb{E}_{N-2} \left[\det\left(M-x_{2K} I \right) \det\left(M-x_{2K-1} I \right)
\prod_{k=1}^{2K-2} s_{x_k}(M) \Lambda^{M}(dx_k)\right].\nonumber
\end{eqnarray}
An application of further $(2K-2)$ Edelman transforms leads
to the desired expression for the modified density:
\begin{eqnarray}
&& \hspace{-.3in} \tilde{\rho}^{(N)}(x_1,x_2,\ldots, x_{2K} )  \nonumber \\
& = & \frac{\Velta(\mathbf{x}) }{16^K} \prod_{k=1}^{2K} \left( |S_{N-k}|
\pi^{-\frac{N-k}{2}} e^{-x_k^2} \right)  \mathbb{E}_{N-K} \! \left[\prod_{m=1}^{2K}\det\left(M-x_m I \right)\right].
\label{app:md}
\end{eqnarray}
Lemma \ref{lm1} is proved with
\[
c_1(N,K)=\prod_{k=1}^{2K} \left( \frac{|S_{N-k}|
\pi^{-\frac{N-k}{2}}}{4} \right).
\]



\subsection{Theorem \ref{thm1}}
\label{ss2}
The full proof of this theorem can be found in the appendix of \cite{roger_oleg_2014},
the main topic of which is the study of Brownian motion taking values in real matrices.
Here we sketch the main steps of the proof.  
The integral representation for the expectation value of a product of characteristic
polynomials can be  derived for any Gaussian random matrix ensemble following
\cite{Brezin-Hikami}, see \cite{sommers2008general} for the specific
case of the real Ginibre ensemble. As a first step, the determinants are represented as Gaussian
integrals over anti-commuting (Grassmann) variables. The integral
with respect to the random matrix measure becomes Gaussian as well and can then be computed
exactly.
This leads to an integral representation for the expectation of a product
of $K$ characteristic polynomials as a Berezin integral with respect to $O(KN)$ variables.
The integrand is the exponential of a polynomial of the fourth degree
in anti-commuting variables. Finally, the Berezin integral can be re-written as a bosonic
integral over $K(K-1)/2$ complex variables using a Hubbard-Stratonovich transformation and
also computed exactly. The answer is
\bea
&& \hspace{-.3in} \mathbb{E}_{N} \left[\prod_{m=1}^K \det \left(M-x_m I\right)\right] \nonumber \\
& = &  \prod_{1\leq p< q \leq K} \left[ \int_{\mathbf{\R^2}} \frac{dz_{pq} d\bar{z}_{pq}}{\pi}
e^{-|z_{pq}|^2} \right] Pf
\left( \begin{array}{cc}
\frac{1}{\sqrt{2}}Z & X  \\
-X & \frac{1}{\sqrt{2}}Z^{\dagger}
\end{array} \right)^N. \label{app:int1}
\eea
Here each $dz_{pq} d\bar{z}_{pq}$ is shorthand for Lebesgue measure on $\R^2$ and arises
from repeated use of the Hubbard-Stratonovich transform; $X=\mbox{Diag}(x)$ with $x \in \R^K$;
and $Z$ is a skew symmetric
complex $K\times K$ matrix.
The right hand side of expression (\ref{app:int1}) can be re-written as a matrix integral:
\bea\label{app:int2}
 \pi^{-\frac{K(K-1)}{2}}
 \int_{Q^{(K)}} \lambda(dZ,dZ^{\dagger}) e^{-\frac{1}{2} Tr ZZ^{\dagger}}
Pf
\left( \begin{array}{cc}
\frac{1}{\sqrt{2}}Z & X  \\
-X & \frac{1}{\sqrt{2}}Z^{\dagger}
\end{array} \right)^N,
\eea
where $Q^{(K)}=\{ Z\in \mathbf{C}^{K\times K}\mid Z^T=-Z\}$ is the space of skew-symmetric
complex matrices and $\lambda (dZ,dZ^{\dagger})$ is the Lebesgue measure on $Q^{(K)}$ as described above.

Note that the dimension of the integral
in the right hand side of (\ref{app:int2}) is $N$-independent. The size of the original
matrix only enters the integral as the power of the Pfaffian in the integrand. This
allows one to calculate the large $N$-limit of (\ref{app:int2}) using the Laplace method.
To facilitate the application of asymptotic methods, one rescales the integration variables
$(Z,\dZ) \rightarrow \sqrt{N} (Z,\dZ)$ to arrive at
\begin{equation}
\label{app:int3}
\mathbb{E}_{N} \left[\prod_{m=1}^K\det\left(M-x_m\right)\right]
 =  \pi^{-\frac{K(K-1)}{2}} 2^{-\frac{NK}{2} }N^{\frac{NK}{2}} N^{\frac{K(K-1)}{2}} J_N
\end{equation}
where
\bea\label{app:io}
J_N = \int_{Q^{(K)}} \lambda(dZ,dZ^{\dagger}) e^{-\frac{N}{2} Tr ZZ^{\dagger}}
Pf
\left( \begin{array}{cc}
Z & \sqrt{\frac{2}{N}}X  \\
-\sqrt{\frac{2}{N}}X & Z^{\dagger}
\end{array} \right)^N \!\!\! \!\!\!.
\eea
The integrand in $J_N$ is now of the form $\exp (NF_N(Z))$, where $F_N$ is a slow function of $N$ in the sense
that $F_N$ and its derivatives converge in the limit $N\rightarrow \infty$.
The main contribution to (\ref{app:io}) for $N\rightarrow \infty$ comes from the neighborhood of the
points of global minimum of the function
\bea
F_{\infty}(Z)=Tr ZZ^{\dagger}-\ln \det\left(Z\dZ\right),~Z \in Q^{(K)}.
\eea
The global minimum value of $F_{\infty}$ is $K$ and it is attained on the set
\begin{equation} \label{CK}
aU(K) = \{W \in Q^{(K)}\mid WW^\dagger=I \}.
\end{equation}
of the skew-symmetric unitary $K \times K$ matrices.

The set $aU(K)$ is a smooth sub-manifold
$Q^{(K)}$. It is  also a non-degenerate critical set meaning that the Hessian of $F_\infty$ has the maximal possible
rank at every point of $aU(K)$. Therefore we can use the standard multi-dimensional Laplace Theorem \cite{erdelyi2010asymptotic} to
calculate the asymptotic expansion of $J_{N}$.  The final answer is
\begin{equation} \label{J0}
J_{N} = e^{- \frac{NK}{2}} (2 \pi N)^{-\frac{K(K-1)}{4}}
\int_{aU(K)}\mu(dW) e^{Tr\left(W^{\dagger}X W X\right)} (1+O(N^{-1})),
\end{equation}
where $\mu(dW)$ is the Haar measure on $aU(K)$ which can be defined as the unique
probability
measure invariant with respect to the following transitive action of $U(K)$ on $aU(K)$:
\bea
U(K)\times aU(K) &\longrightarrow& aU(K)\\
(U,W) &\mapsto& UWU^T.\nonumber
\eea

Collecting together (\ref{app:md}), (\ref{app:int3}) and (\ref{J0}) we find
\bea
\tilde{\rho}^{(N)}(x_1,x_2,\ldots, x_K )  = c_2(N,K)
\Velta(\mathbf{x}) \prod_{k=1}^{K} e^{-x_k^2} \int_{aU(K)}\mu(dW) e^{Tr\left(W^{\dagger}X W X\right)}
\left(1+ o(1)\right),\nonumber\\
\eea
where
\begin{eqnarray}
c_2(N,K) & = &  C_K  \prod_{k=1}^{K} \left( |S_{N-k}|  \pi^{-\frac{N-k}{2}} \right)
\pi^{-\frac{K(K-1)}{2}} 2^{-\frac{(N-K)K}{2} } \\
&& \hspace{.2in} (N-K)^{\frac{(N-K)K}{2}}
(N-K)^{\frac{K(K-1)}{2}}
e^{- \frac{(N-K)K}{2}} (2 \pi (N-K))^{-\frac{K(K-1)}{4}}\nonumber
\end{eqnarray}
and $C_K>0$ denotes a $K$-dependent constant. It is lengthy but straightforward
to check that $c_2(N,K) \to c_3(K) >0$ as $N \to \infty$
and hence that the limiting modified density $\tilde{\rho}(x_1,x_2,\ldots, x_K ) := \lim_{N \to \infty}
\tilde{\rho}^{(N)}(x_1,x_2,\ldots, x_K ) $ exists and is given by
\begin{equation} \label{temp11}
\tilde{\rho}(x_1,x_2,\ldots, x_K )  = c_3(K)
\Velta(\mathbf{x}) \prod_{k=1}^{K} e^{-x_k^2} \int_{aU(K)}\mu(dW) e^{Tr\left(W^{\dagger}X W X\right)}.
\end{equation}
By the spectral theorem for skew-symmetric unitary
matrices, $W\in aU(K)$ iff there is $U\in U(K)$
such that $W=UJU^T$, \cite{mehta1989matrix}.
The pullback of the Haar measure on $aU(K)$is a Haar measure on $U(K)$
under the map $U\rightarrow W$.
Under this map the integral (\ref{temp11})
coincides with (\ref{lm2}).
Theorem \ref{thm1} is proved.
\subsection{Proposition \ref{lm2}}
\label{ss3}
\subsubsection{Equation}
Let us start with a general observation concerning solutions to the heat equation
on linear spaces.
Consider the canonical heat equation on a real $n$-dimensional vector space $\V$, equipped with
an inner product $\langle \cdot,\cdot\rangle$:
\bea\label{ghe}
\left(\frac{\partial}{\partial t}- \frac{1}{2}\mbox{div}~ \mbox{grad}\right) \phi_t(x)=0, ~\lim_{x \rightarrow \infty}
\phi_t(x)=0.
\eea
Using explicit coordinates on $\V$, 
$\mbox{div}~\mbox{grad}=\sum_{i,j=1}^n g^{ij}\frac{\partial}{\partial x_i}\frac{\partial}{\partial x_j}$, 
where $(g^{ij})_{1\leq i,j\leq n}$ is the inverse of the matrix
defining the inner product.
The fundamental solution is
\bea
\Phi_t(x)=\frac{1}{(2\pi t)^{n/2}}e^{-\frac{\langle x, x \rangle}{2t}}.
\eea

Let $P:\V\rightarrow \V$ be an orthogonal projection operator, that is $P^2=P$, $P=P^T$.
Then it is straightforward to check that
\bea\label{fs1}
\Phi_t(x \mid P)=\frac{1}{(2\pi t)^{rank(P)/2}}e^{-\frac{\langle x,P x\rangle}{2t}},
\eea
also solves (\ref{ghe}). This solution can be regarded as fundamental in the space
of initial conditions constant on $Ker(P)$.

Let $\{P_g\}_{g \in \G}$ be a (continuous) family of projection operators 
parameterized by points of a compact measure space $\G$ with a finite measure $\mu$.
Then, by (\ref{fs1}) and the linearity of the heat equation, the function
\bea
\Phi_t(x \mid P,\mu)=\int_{\G}\frac{\mu(dg)}{(2\pi t)^{rank(P_g)/2}} e^{-\frac{\langle x,P_g x \rangle}{2t}},
\eea
also solves (\ref{ghe}).

Now consider an integral representation for the 
modified density
\bea \label{Irho}
\tilde{\rho}_t(X):=C_K\left(\sqrt{t}\right)^{-\frac{K(K+1)}{2}}\Velta(X)I_t(X),
\eea
where $X=\mbox{Diag}(x)$ for $x \in \R^K$ and
\bea\label{intl1}
I_t(X)=\prod_{k=1}^{K} e^{-x_k^2/t} \int_{aU(K)}\mu(dW) e^{\frac{1}{t} Tr\left(W^{\dagger}X W X\right)}.
\eea
The total power of $\sqrt{t}$ corresponds to the diffusive rescaling of $\tilde{\rho}(x)\prod_{k=1}^{K}dx_k$
from (\ref{rho_mod}). By changing variables in the above integral, $W\rightarrow UWU^T$, where
$U$ is a fixed unitary matrix, 
\bea\label{intl2}
I_t(X)=\int_{aU(K)}\mu(dW) e^{-\frac{1}{t} Tr\left(-W^{\dagger}H W H^T+H^2\right)},
\eea
where $H=UXU^\dagger$ is a Hermitian matrix with eigenvalues given by
$X$. 
The integral (\ref{intl2}) can be further re-written as
\bea
I_t(H)=\int_{aU(K)}\mu(dW) 
e^{-\frac{1}{t}Tr\left(H P_W (H)\right)},
\eea
where
\bea
P_W=I+W\otimes W^\dagger\circ\hat{T}
\eea
is a linear operator acting in the space of $K\times K$ Hermitian
matrices, $\hat{T}$ is the operator of transposition; the action of
the linear operator $A\otimes B$ is defined according to the formula
\[
A\otimes B(C)=ACB^T
\]
The operator $P_W$ is proportional to a projector operator:
for any Hermitian matrix $H$,
\bea
P_W^2(H)=H+2W\otimes W^\dagger \circ \hat{T}(H)+
W(WH^TW^\dagger)^TW^\dagger=2P_W(H),
\eea
where we used $WW^\dagger =I$ and $W^T=-W$. So $\frac{P_W}{2}$ is a projection
operator, which is also self-adjoint, and we can apply the theory outlined in the beginning of the subsection by
identifying $\V$ with the real vector space of $K\times K$ Hermitian matrices with the inner product given
by $(A,B)\mapsto Tr AB$.
As
\[
rank\left(\frac{P_W}{2}\right)=\frac{1}{2} Tr P_W=\frac{1}{2}(K^2+K),
\]
we can conclude that 
\bea
\left(\frac{\partial}{\partial t}-\frac{1}{8}\Delta_H\right) t^{-\frac{K(K+1)}{4}}I_t(H)=0
\eea
Recall that $I_t(H)=I_t(UXU^{\dagger})=I_t(X)$ due to the $U(K)$-invariance.
Therefore, the above equation reduces to the 'radial' form well used in
random matrix theory,
\bea\label{intl5}
\left(\frac{\partial}{\partial t}-\frac{1}{8}\sum_{k=1}^K
\frac{1}{\Velta^2(x)}\frac{\partial}{\partial x_k}\Velta^2(x)
\frac{\partial}{\partial x_k}\right) t^{-\frac{K(K+1)}{4}} I_t(X)=0.
\eea
The same equation is satisfied by the Itzykson-Zuber integral albeit for a 
different choice of initial conditions.
Using this equation for $I_t$ in (\ref{Irho}) we find the modified
density $\tilde{\rho}$ satisfies the canonical 'flat' heat equation:
\bea\label{intl7}
\left(\frac{\partial}{\partial t}-\frac{1}{8}\sum_{k=1}^K
\frac{\partial^2}{\partial x_k^2}\right) \tilde{\rho}_t(X)=0, ~t>0, x\in \R^K.
\eea


\subsubsection{Initial conditions}
\label{ss4}
Without loss of generality, due to the permutation symmetry of $\tilde{\rho}_t$,
we can assume that $x \in W^{(K)}$. 
The initial condition for (\ref{intl7}) follows from the asymptotic analysis
of the integral (\ref{intl1}) in the limit $t\downarrow 0$. Our aim is to prove that
\bea\label{ic5}
\tilde{\rho}_{0+}(x)= C_K \prod_{k=1}^{K/2}\delta^\prime(x_{2k}-x_{2k-1}),
\eea
where $\delta^{\prime}$ is the derivative of the Dirac's delta function understood
in the distributional sense. The proof should be read after the proof of Theorem \ref{thm2}
in Section \ref{ss5}, where all the relevant notions used in the current section are defined.

\begin{figure}[htb]\label{fig111}
\begin{center}
\includegraphics[height=4in,width=5in,angle=0]{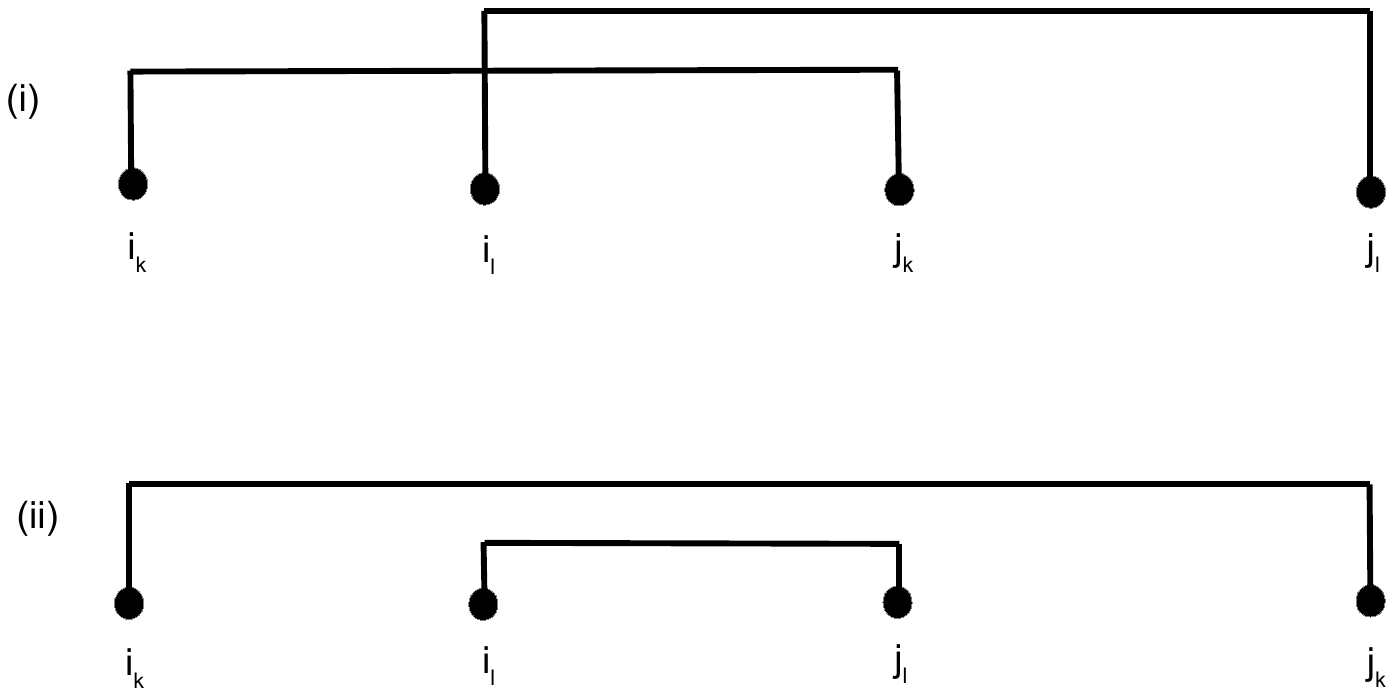}
\caption{Two possible matchings with $j_k>i_l$ for $k<l$: (i) single crossing; (ii) double crossing.}
\end{center}
\end{figure}

The set of the critical points for the integral in (\ref{intl1}) is given by (\ref{setofcp}), which is valid for any
$t \in \C\setminus\{0\}$. But, unlike the imaginary case, the asymptotics for $t>0$
is determined by
the global maximum of the function $F_X: aU(K)\rightarrow \R$ given by
$F_X(W) = Tr(W^{\dagger}XWX)$.
Note that $F_X$ is constant on each torus $\T_{\sigma}$, that is it does not depend on the 
phase $\Phi$. 
We claim that the global maximum of $F_X$ is achieved on $\T_{\pi_0}$ for 
the matching $\pi_0=(1,2)(3,4)\ldots(K-1,K)$.
Let us prove this
claim by contradiction. 
Assume that the global maximum is reached at a matching $\pi_1\neq \pi_0$,
\[
\pi_1=(i_1, j_1)(i_2,j_2)\ldots (i_{K/2},j_{K/2}),
\]
where $i_k<i_l$ for $k<l$ and $i_k<j_k$ for any $k,l$ between $1$ and $K/2$.
Notice that using the $(i,j)$ notations, the matching $\pi_0$ can be uniquely 
characterised by the following property: $j_k<i_l$ for any $k<l$.
As $\pi_1\neq \pi_0$, there exists $k<l$ such that $j_k>i_l$, see Fig. \ref{fig111} for the 
illustration of the corresponding matchings. 
Then
\[
F_X(P_{\pi_1})=F_0+2(x_{i_l} x_{j_l}+x_{i_k}x_{j_k}),
\]
where $F_{0}$ is the part of $F_X$ which does not depend on $x_{i_l}, x_{j_l},x_{k_k},x_{j_k}$.
Let $\tilde{\pi}_1$ be a matching obtained from $\pi_1$ by rematching indices $i_k,j_k,i_l,j_l$ as follows:
$(i_k,j_k)(i_l,j_l)\rightarrow (i_l,i_k)(j_k,j_l)$. Then
\[
F_X(P_{\tilde{\pi}_1})=F_0+2(x_{i_l} x_{i_k}+x_{j_l}x_{j_k}),
\]
and
\bea\label{contra}
F_X(P_{\tilde{\pi}_1})-F_X(P_{\pi_1})=(x_{i_l}-x_{j_k})(x_{i_k}-x_{j_l})>0
\eea
due to the ordering of $x$'s, $x_m>x_n$ for $m>n$. 
Here we used the assumed inequality $j_k>i_l$ and the observation $i_k<i_l<j_l$.
The inequality (\ref{contra}) contradicts
our assumption that the global maximum of $F_X$ is achieved at $\pi_1$. We conclude that the global maximum
of $F_X$ is achieved on the critical manifold $\T_{\pi_0}$ where it is constant with
the value 
\bea\label{max}
F_{max}= 2 \sum_{i=1}^{K/2} x_{2k-1} x_{2k}.
\eea
The principal asymptotics of the integral (\ref{intl1}) is given by the integral of the exponent of the order second expansion of
the function $F$  in the vicinity of $\T_{\pi_0}$ over the total space of the normal bundle $N(\T_{\pi_0})$. The fiber
of the normal bundle at $W \in \T_{\pi_0}$ is defined by the orthogonal decomposition $T_W(aU(K))=T_W(\T_{\pi_0})
\oplus N_W(\T_{\pi_0})$ using the Hermitian inner product $tr(A^{\dagger}B)$ inherited from the space of all skew-symmetric
complex matrices. This part of the proof is completely analogous to the computation carried out in Section \ref{ss5}.

Expanding the exponent $F$ near a point of the global maximum $W_c=P_{\pi_0}e^i\Phi$, we find
\bea\label{qf}
F_X(P_{\pi_0}+\xi)=F_{max}+\frac{1}{2}\sum_{m,n=1}^K |\xi_{mn}|^2 
(X_m-X_{\pi_0(n)})(X_n-X_{\pi_0(m)})+O(\xi^3).
\eea
The above quadratic form is degenerate due to the degeneracy
of the critical manifold $\T_{\pi_0}$. It follows from the characterisation (\ref{ts}) of the tangent space
at $W_c$ that for $\pi_0=(12)(24)\ldots (K-1,K)$, the normal space $N_{W_{c}}(\T_{\pi_0})$
is parameterised by complex co-ordinates $$(\xi_{2k-1,2l-1},\xi_{2k-1,2l}))_{1\leq k<l\leq K/2}.$$
In terms of these co-ordinates, the restriction of the quadratic form (\ref{qf}) to $N_{W_{c}}(T_{\pi_0})$ is negatively definite:
\bea\label{qfn}
F_X(P_{\pi_0}+\xi)=F_{max}+2\sum_{1\leq k<l<K/2} |\xi_{2k-1,2l-1}|^2 (X_{2k-1}-X_{2l})(X_{2l-1}-X_{2k})\nonumber\\
+2\sum_{1\leq k<l<K/2} |\xi_{2k-1,2l}|^2 (X_{2k-1}-X_{2l-1})(X_{2l}-X_{2k})+O(\xi^3).
\eea
In the basis described above, the matrix of second derivatives of $F$ evaluated at the critical point
is diagonal. The square root of its determinant is equal to 
\[
\Velta(X)\left(\prod_{k=1}^{K/2}(x_{x_{2k}}-x_{2k-1})\right)^{-1}
\]
We conclude that
\bea\label{answer}
I_t(X)\stackrel{t\rightarrow 0}{\sim} C_{K} \frac{\prod_{k=1}^{K/2}(x_{2k}-x_{2k-1})}{\Velta(X)}
e^{+\frac{2}{t}\sum_{k=1}^{K/2} x_{2k-1}x_{2k}} e^{-\frac{1}{t}\sum_{k=1}^{K} x_{k}^2} ,
\eea
where $C_K$ is a non-zero constant. Substituting (\ref{answer}) into (\ref{Irho})
one finds that the $t\downarrow 0$ limit of  exists in the distributional sense and is equal to
\bea
\tilde{\rho}_{0+}(x)=C_K \prod_{k=1}^{K/2} \delta^{\prime}(x_{2k}-x_{2k-1}).
\eea
\subsection{Theorem \ref{bsfn}}\label{ss4}
Using the one-dimensional heat kernel 
\bea
g_t(x)=\frac{1}{\sqrt{\pi t/2}}e^{-\frac{2}{t}x^2}
\eea
the equation (\ref{ivp}) has the unique solution
\bea
\tilde{\rho}_t(y) = C_K\int_{\R^K} dx_1 dx_2 \ldots dx_K
\prod_{k=1}^K g_t(y_k-x_k) \prod_{k=1}^{K/2}\delta^{\prime}(x_{2k}-x_{2k-1}).
\eea
 For any permutation $\sigma \in \Sigma_{K}$ we have, due to the Vandermonde
 $V(y)$ in its definition, 
 \bea
 \tilde{\rho}_t(y_{\sigma_1}, \ldots, y_{\sigma_K}) = \mbox{sign}(\sigma)   \tilde{\rho}_t(y).
\eea
Thus
\begin{eqnarray*}
\tilde{\rho}_t(y) &=& \frac{C_K}{K!}  \sum_{\sigma \in \Sigma_{K}} \mbox{sign}(\sigma) \int_{\R^K} dx_1 dx_2\ldots dx_K 
\prod_{k=1}^K g_t(y_{\sigma_k} -x_k) \prod_{k=1}^{K/2}\delta^{\prime}(x_{2k}-x_{2k-1})\\
&= &  \frac{C_K}{K!}  \sum_{\sigma \in \Sigma_{K}} \mbox{sign}(\sigma) \prod_{k=1}^{K/2} 
\int_{\R^2} g_t(y_{\sigma_{2k}} -x) g_t(y_{\sigma_{2k-1}} -x') \delta^{\prime}(x-x') dx dx'\\
&=& \frac{C_K}{K!} \pf_{1\leq i,j\leq K}\left[
\frac{\partial}{\partial y_i}g_{2t}(y_i-y_j)\right].
\end{eqnarray*}
%
%
The derivation of the Pfaffian here is analogous to that in the de Bruijn formula
\cite{de1955some}. 

Specializing to $t=1$, we conclude for $x_1< \ldots < x_K$ that
\begin{equation} \label{temp12}
\tilde{\rho}(x_1,x_2,\ldots, x_K )  = C_K \pf \left[(x_i-x_j)e^{-(x_i-x_j)^2}: 1 \leq i < j \leq K \right].
\end{equation}
The correlation functions of spins can be formally computed by integrating $\tilde{\rho}$ with respect to space variables:
for $x_1< \ldots < x_K$ 
\begin{eqnarray}\label{eq:ss1d}
\mathbb{E}\left[\prod_{k=1}^K s_{x_k}(M) \right] = (-2)^K\left(\prod_{k=1}^K \int_{-\infty}^{x_k} dy_k\right)
\tilde{\rho}(y_1,y_2,\ldots, y_{k}).
\end{eqnarray}
This leads to the spin-spin correlation function: for $x_1< \ldots < x_K$ 
\begin{equation} \label{temp13}
\mathbb{E} \left[\prod_{k=1}^K s_{x_k}(M) \right]=
C_K  \pf \left[ \int^{\infty}_{x_j-x_i} e^{-z^2} dz : 1 \leq i < j \leq K \right],
\end{equation}
which coincides with the correlation function of spins in the continuous
limit of the kinetic Glauber spin chain, which justifies our choice of
terminology. 
The constants $C_k$ can be found inductively in $k$ by allowing $x_{2k} \downarrow x_{2k-1}$, and
noting that $\E[s_{x_1}(M)s_{x_1}(M)] =1$. This yields $C_K = (4/\pi)^{K/4}$.

Finally, substituting (\ref{temp13}) into formula (\ref{eq:dens})
and performing the differentiation explicitly, we find
\bea
\rho(x)=\pf_{1\leq i,j\leq K}\big[H(x_j-x_i) \big],
\eea
where 
\bea
H(z) = \left( \begin{array}{cc}
 -F''(z) & -F'(z) \\
F'(z) & \sgn(z) F(|z|)
\end{array} \right)
\label{oldclaim2}
\eea
and
\bea
F(z) = \pi^{-1/2} \int^{\infty}_z e^{-x^2} dx.
\label{errf}
\eea
Here $\sgn(z) = 1$ for $z>0$, $\sgn(z) = -1$ for $z<0$ and $\sgn(0) =0$.

\subsection{Theorem \ref{thm2}}\label{ss5}
Our aim is to calculate the leading term in the small-$t$ asymptotic of 
\bea\label{intl9}
I_{it}(X)=   \prod_{k=1}^{2K} e^{-\frac{1}{it}x_k^2}  \prod_{k=1}^{2K} \int_{aU(2K)}
\mu(dW) e^{+\frac{1}{it}F_X(W)},
\eea
where $F_X(W)=Tr(W^{\dagger}XWX)$ and $X=\mbox{Diag}(x)$ with $x \in \R^{2K}$ and  $x_i\neq x_j$
for $i \neq j$. Due to the permutation invariance of $F_X$, we may assume that
$x_i<x_j$ for $i<j$. The small-$t$ asymptotics
of the integral (\ref{intl9}) can be found by adapting the standard
multi-dimensional stationary phase method \cite{st_phase} as follows.

The main contribution to the integral for $t\rightarrow 0$ comes from the vicinity of 
the critical points of the function $F_X: aU(2K)\rightarrow \R$. 
Its calculation requires the knowledge of the set of the critical points and
the expansion of $F_X$ around this set.

To derive
the critical point equation we need a parameterisation of the tangent space
$T_W(aU(2K))$ at $W\in aU(2K)$:
\bea\label{ts}
T_W(aU(2K))=\{\xi \in \mathbb{C}^{4K^2}\mid \xi^T=-\xi; W\xi^\dagger W=-\xi\}.
\eea
In other words, the tangent space is spanned by skew-symmetric, $W$-anti
self-dual $2K\times 2K$ complex matrices, where $W$-dual of a complex matrix $\alpha$
is $\alpha^R:=W\alpha^\dagger W.$ 

The critical point condition is the vanishing of the directional derivative 
of $F_X$, $D_\xi F_X(W)=0$. Explicitly this leads to
\[
[X,WXW^{\dagger}]=0.
\]
As $X$ is diagonal with distinct diagonal entries, the vanishing of the commutator
$[X,WXW^{\dagger}]$ means that the matrix $WXW^\dagger$
is diagonal,
\[
WXW^\dagger=D.
\]
As $D$ is similar to $X$, its diagonal entries are a permutation of the 
diagonal entries of $X$. Therefore, $W_0$ is a critical point of $F_X$ if
\[
W_0=P_\sigma e^{i\Phi},
\]
where $P_\sigma$ is the permutation matrix corresponding to the permutation 
$\sigma$, an element of the permutation group $S(2K)$,
and
$\Phi$ is a diagonal real matrix.
As it is easy to check, the skew-symmetry condition, $W=-W^T$ implies
that: 
\begin{enumerate}
\item The permutation $\sigma$ is a product of two-cycles, meaning that $(P_\sigma)_{ii}=0$
for any $i=1,2,\ldots, 2K$ and $P_\sigma=P_\sigma^T$; the elements of the set $M(2K)\subset S(2K)$ satisfying these are called matchings.
\item $\Phi_{\sigma(i)\sigma(i)}=\Phi_{ii}+\pi$
for any $i=1,2,\ldots, 2K$.
\end{enumerate}

We conclude that the set of the critical points of the function $F_X$ is
\begin{eqnarray}
\mathbf{C}=\cup_{\sigma \in M(2K)} \T_\sigma,
\label{setofcp}
\end{eqnarray}
where 
$$\T_{\sigma}=\{ P_\sigma e^{i\Phi}; e^{i\Phi} \in U(1)^K: \Phi_{\sigma(i)\sigma(i)}=\Phi_{ii}+\pi\}\subset aU(2K).$$
In other words,
$\mathbf{C}$ is a union of non-intersecting $K$-dimensional tori. 
The restriction of the exponent $F_X$ to the critical manifold $\T_{\sigma}$ is a constant equal to
\bea\label{fc}
F_X(\sigma)=\sum_{i=1}^{2K} x_{\sigma(i)} x_i.
\eea
The second order Taylor expansion of $F_X$ in the vicinity of the critical manifold is given,
for $W(\sigma, \Phi)+\xi \in aU(2K)$, by
\bea\label{qf}
F_X(W(\sigma, \Phi)\!+\!\xi)\!=\!F_X(\sigma)+\frac{1}{2}\!\!\sum_{m,n=1}^{2K}
|\xi_{mn}|^2 (x_m\!-\!x_{\sigma(n)})(x_n\!-\!x_{\sigma(m)})\!+\!O(\xi^3).
\eea
The quadratic form defining the second order term is constant on $\T_\sigma$.
It is also degenerate as the set of the critical
points 
$\T_{\sigma}$ is not isolated. A calculation shows that 
$F_X(W(\sigma, \Phi)+\xi)=F_X(\sigma)+
O(\xi^3)$ if $\xi \in T_{W}(\T_{\sigma})\subset T_W(aU(2K))$ for any $W\in \T_\sigma$.
The critical manifold is non-degenerate in the sense that the quadratic form entering
(\ref{qf}) has the maximal rank, see below. 

\begin{figure}[htb]\label{fig112}
\begin{center}
\includegraphics[height=3in,width=4.5in,angle=0]{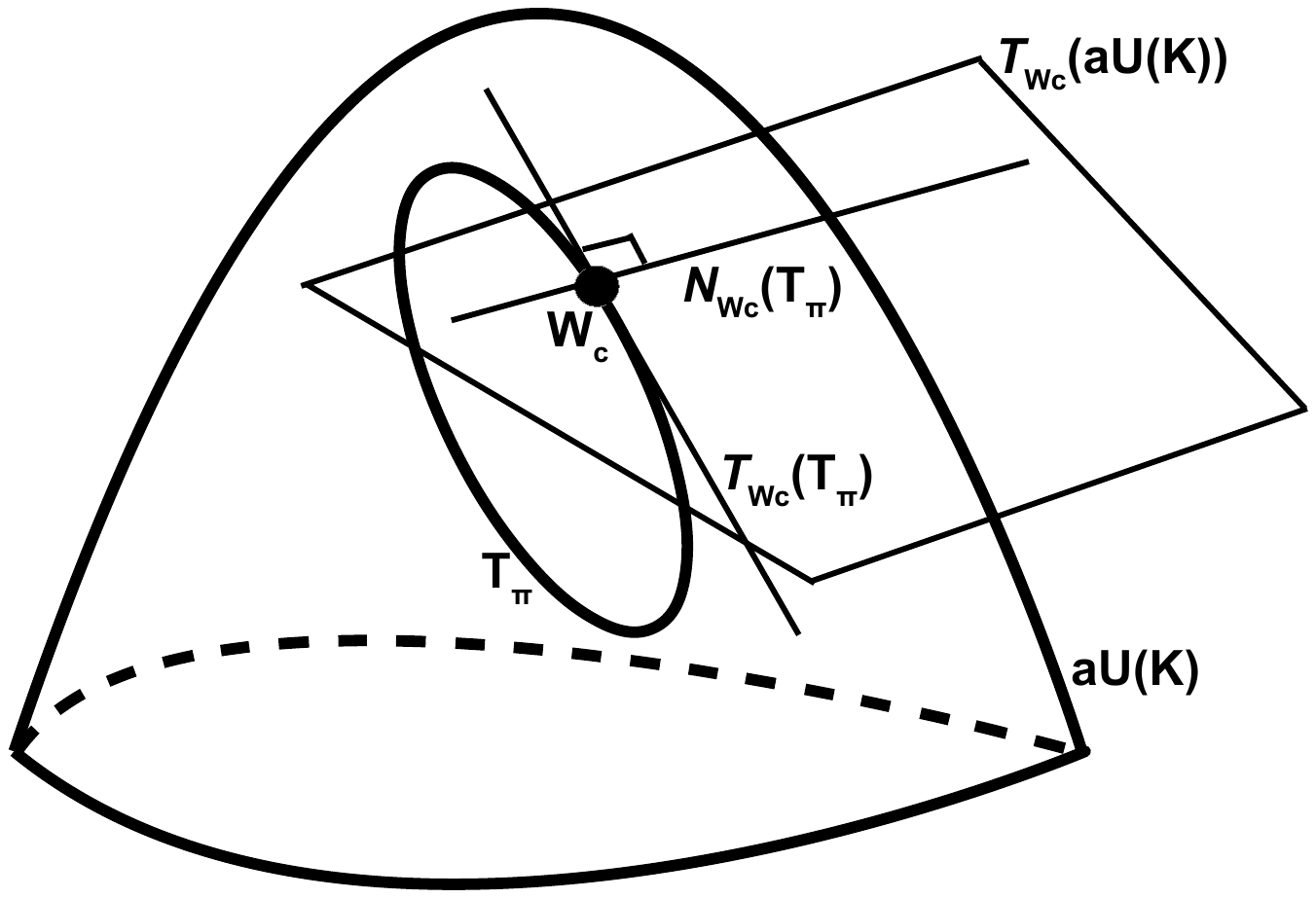}
\caption{The sketch of the integration space for the asymptotic evaluation of the integral (\ref{intl9}).}
\end{center}
\end{figure}

The small-$t$ asymptotic of the integral $I_{it}(X)$ is given by the sum of contributions from each of the
tori $\T_\sigma$. Each contribution is equal to the volume of $\T_\sigma$ multiplied
by a Gaussian integral over the normal space to $\T_\sigma$ at any point on the torus,
say $P_\sigma$. The normal space $N_{\sigma}(\T_{\sigma})$ is
defined by the orthogonal decomposition $T_\sigma(aU(K))=\T_\sigma(T_{\pi_0})
\oplus N_\sigma(\T_{\sigma})$ obtained using the Hermitian inner product $Tr AB$ inherited from the space of all skew-symmetric
complex matrices. The volumes of different tori are equal due to the $U(2K)$-invariance.
The geometry of the integration space is illustrated
in  Fig. \ref{fig112}.    

As the problem is reduced to integrals over linear spaces,
standard stationary phase formulae apply, see for example \cite{st_phase}, Chapter III.
So, using the results collected above, we conclude that
\bea\label{intlstep1}
I_{it} = C_K t^{K(K-1)} \prod_{k=1}^{2K} e^{-\frac{1}{it}x_k^2} \sum_{\sigma \in M(2K)}
\frac{e^{+\frac{1}{it}\sum_{i=1}^{2K} x_{\sigma(i)} x_i+\frac{i\pi}{4}\mbox{sig}(\partial\otimes\partial F_X(\sigma))])}}
{\sqrt{|\det[\partial\otimes\partial F_X(\sigma))]|}}(1+O(t^{\frac{1}{2}})),
\eea
where $\partial\otimes\partial F_X(\sigma)$ is the restriction of the Hessian of $F_X$ at $\sigma$ to 
the normal space, and $\mbox{sig}$ is its associated siganture, 
both of which we will compute using (\ref{qf}). 
The power of $\sqrt{t}$ is equal to the dimension of the normal space which is in turn
equal to $\dim aU(2K)-\dim \T_\sigma=K(2K-1)-K=2K(K-1)$. 
The final step is the calculation of the determinant
and the signature of the Hessian. 

The expression for $\sigma \in M(2K)$ in cycle notation is
\[
\sigma=(i_1 j_1)(i_2 j_2)\ldots (i_k j_k),
\]
where $i_k<j_k$, $1\leq k \leq K$ and $i_k<i_l$ for $1\leq k<l\leq K$.
Using the characterization (\ref{ts}) of the tangent space
at $P_\sigma$, the normal space $N_{P_{\sigma}}(\T_{\sigma})$
can be parameterised by complex co-ordinates $$(\xi_{i_k,i_l},\xi_{i_k,j_l}))_{1\leq k<l\leq K}.$$
The restriction of the quadratic form (\ref{qf}) to $N_{P_\sigma}(\T_{\sigma})$ takes the form
\bea\label{qfn}
F_X(P_\sigma+\xi)-F_X(\sigma)=2\sum_{1\leq k<l\leq K} |\xi_{i_k,i_l}|^2 (x_{i_k}-x_{j_l})(x_{i_l}-x_{j_k})\nonumber\\
+2\sum_{1\leq k<l\leq K} |\xi_{i_k,j_l}|^2 (x_{i_k}-x_{i_l})(x_{j_l}-x_{j_k})+O(\xi^3).
\eea
In the basis described above, the Hessian $\partial \otimes \partial F_X(\sigma)$ 
is diagonal.
The square root of the modulus of its determinant is equal to 
\bea\label{det}
2^{K(K-1)}\big|\prod_{k<l}(x_{j_l}-x_{i_k})(x_{i_l}-x_{j_k}) (x_{i_l}-x_{i_k})(x_{j_l}-x_{j_k})\big|
=
\frac{2^{2K(K-1)}\Velta(X)}{\prod_{k=1}^{K}(x_{j_k}-x_{i_k})},
\eea
where the ordering $x_1<x_2<\ldots <x_{2K}$ can be used to show that the r.h.s. is positive.
To calculate the signature of the Hessian, notice that each summand in (\ref{qfn})
corresponds to an eigenvalue with multiplicity two due to complexity of $\xi$'s. 
The eigenvalue $(x_{i_k}-x_{j_l})(x_{i_l}-x_{j_k})$, where $k<l$ 
is positive if $j_k>i_l$. The eigenvalue $(x_{i_k}-x_{i_l})(x_{j_l}-x_{j_k})$, where
$k<l$ is positive if $j_k>j_l$. In each of these cases the eigenvalue is positive
if the permutation $\pi(\sigma):=(1,2,\ldots,2K)\rightarrow \sigma=(i_1,j_1)(i_2,j_2)\ldots(i_K,j_K)$
has an inversion. Let us stress that here we are treating $\sigma$ as a matching, not as a
permutation.

We conclude that the total number of positive eigenvalues of the Hessian
is equal to twice the total number of inversions in the permutation $\pi(\sigma)$,
which we denote by
$\mbox{inv}(\pi(\sigma))$. 
Then
\bea\label{sign}
\mbox{sign}(\partial\otimes \partial)F_X(\sigma)&=&
\#\mbox{pos. eigenvalues}-\#\mbox{neg. eigenvalues}
\nonumber\\
&=&2~\#\mbox{pos. eigenvalues}-2K(K-1)
\nonumber\\
&=&4~\mbox{inv}(\sigma)-2K(K-1),
\eea
where $2K(K-1)$ is the dimension of the normal space. 
Finally, let us notice that $\exp(i\pi \mbox{inv}(\pi(\sigma))))=\mbox{sign}(\pi(\sigma))$,
the sign of the permutation $\pi(\sigma)$.
Using this observation and
substituting (\ref{det}), (\ref{sign}) into (\ref{intlstep1})
one finds
\bea\label{intlstep2}
I_{it}&=&C_K t^{K(K-1)} \prod_{k=1}^{2K} e^{-\frac{1}{it}x_k^2}  
\sum_{\sigma \in M(2K)}\mbox{sign}(\pi(\sigma))
\frac{\prod_{k=1}^{K}(x_{j_k}-x_{i_k})e^{+\frac{2}{it} x_{i_k} x_{j_K}}}
{\Velta(X)}(1+O(t^{\frac{1}{2}}))\nonumber
\\
&=&C_K\frac{\pf_{1\leq i<j\leq 2K}[\frac{(x_j-x_i)}{\sqrt{t}}\exp[-\frac{1}{it}(x_i-x_j)^2]]}{\Velta(X/\sqrt{t})}(1+O(\sqrt{t})),
\eea
where the definition of the Pfaffian was used in the last step.
Comparing this answer with the exact result (\ref{iz12}), we conclude that the untracked constant $C_K$ 
must agree with the exact answer for all $K \geq 1$
and that the error term $O(\sqrt{t})$ in this stationary phase expansion is in fact identically zero. The asymptotic exactness is proved.



\vspace{.1in}

\noindent
{\bf Acknowledgements.} Our research was partially supported by EPSRC and the Leverhulme
trust. We are grateful to Yan Fyodorov and Gernot Akemann for many inspiring discussions. 

\end{document}